\documentclass[showpacs,preprintnumbers,superscriptaddress]{revtex4}

\usepackage{amsmath,amssymb,graphicx,bm}

\begin{document}
%\begin{CJK*}
%\preprint{APS/123-QED}

\title{Asymptotically safe phantom cosmology}

\author{Rong-Jia Yang}
\email{yangrongjia@tsinghua.org.cn}
 \affiliation{College of Physical Science and Technology, Hebei University, Baoding 071002, China}
 \affiliation{Department of Physics, Tsinghua University, Beijing 100084, China}

\date{\today}

\begin{abstract}
We consider quantum modifications to phantom cosmology in a Friedmann每Robertson每Walker spacetime. The cosmological evolution equations improved by the renormalization group are obtained. For exponential potential, we find two types of cosmological fixed point; the renormalization group scale either freezes in, or continues to evolve with scale factor. We discuss the implications of each of these points, and investigate especially whether the big rip can be avoided. If the fixed point of renormalization group flow coincides with the cosmological fixed point, the universe will be dominated by dark matter and will be free from the big rip.

\end{abstract}

\maketitle
%\end{CJK*}

\section{Introduction}
The simplest and most theoretically appealing candidate of
dark energy is the vacuum energy (or the cosmological constant
$\Lambda$) with a constant equation of state (EoS) parameter $w=-1$.
This scenario is in general agreement with the current astronomical
observations, but has difficulties to reconcile the small
observational value of dark energy density with estimates from
quantum field theories; this is the cosmological constant problem.
Recently it was shown that $\Lambda$CDM model may also suffer from the age problem \cite{Yang2010}.
It is thus natural to pursue alternative possibilities to explain
the mystery of dark energy. Over the past decade numerous dark energy models have been proposed,
such as quintessence, phantom, k-essence, tachyon, (Generalized) Chaplygin Gas, DGP, etc. The phantom, with $w_\phi<-1$, has received much attention recently \cite{Caldwell}. It has some strange properties. For example, it violates the dominant-energy condition \cite{Carroll2003}, which will result in instability \cite{Carroll2003,Carroll2005}. More uncomfortably, its energy density increases with time, which will lead to cosmic doomsday in finite time \cite{Caldwell03}. Despite these shortcomings, however, we can reasonably conceive that at low scales new physics in the dark-energy sector emerges to render a phantom model stable, or that a mechanism can mimic phantom behavior (see, for example, modifications of the Friedmann equation), as discussed in \cite{Carroll2003}; on the other hand, observations 
cannot rule out the possibility $w<-1$ \cite{Melchiorri2003, Gaztanaga2009}, and given how little we understand about dark energy, we should keep an open mind about the discussions of phantom. Besides the stability, how to avoid the big rip in phantom cosmology is also worthwhile to deeply investigate, which is the topic here. One possible way is to take into account effects coming from quantum gravity, which may affect the behavior of the phantom.

The renormalization group (RG) is an efficient way of analyzing quantum effects on the low energy scale \cite{Wilson}. This technique allows to derive equations for the running of couplings of an effective average action without any power expansion in the couplings. It has been conjectured that the renormalization group flow of the couplings of the theory would have a non-trivial fixed point, with a finite dimensional ultraviolet critical surface of trajectories attracted to the fixed point in the UV scale limit. Since this running would counteract the divergencies, it is called an asymptotic safety scenario \cite{Weinberg}, for reviews see \cite{Litim, Niedermaier, Percacci}, and references therein. This exciting idea has been applied to investigate the existence of the UV fixed point in Einstein gravity \cite{Bonanno}, scalar-tensor theory \cite{Reuter2004,Cai}, $f(R)$ gravity \cite{Benedetti,Souma,Lauscher,Reuter2002,Litim2008,Machado,Codello}, and so on.
Recently, It also has been postulated that an IR fixed point in quantum gravity exists, which has been explored in cosmology \cite{Bonanno1}. However, the possibility of an IR fixed point in quantum gravity is still a speculative idea, and there is as yet no direct evidence from any computed $\beta$-functions. References \cite{Hindmarsh,Linder,Hindmarsh1} have adapted cosmological RG ideas to include quintessence scalar field and discussed its effects on the cosmological behavior.
Assumption that the RG scale is proportional to $1/t$ has been explored in \cite{Bonanno,Bonanno1,Reuter}, while in \cite{Shapiro2002} the scale $k\sim H$ has been taken. Other choices link the RG scale with the cosmological event \cite{Bauer}, or the fourth root of the energy density \cite{Shapiro2000} on which Ref. \cite{Guberina} based. Perturbative RG approaches have also been implicated for deviations from standard cosmology \cite{Shapiro} or black hole \cite{Casadio}. A more general scale-setting procedure for General Relativity with renormalization Group corrections was proposed in \cite{Domazet}. Ref. \cite{Rodrigues} considered the application of quantum corrections computed using renormalization group arguments in the astrophysical domain. Here we focus on the IR behavior of the field and its effects on phantom dark energy and the cosmological expansion. We will look for cosmological fixed points to the coupled dynamical equations including RG effects, and investigate whether the big rip can be avoided when the potential varies as the RG scale $k$ changes.

This paper is organized as follows, in the following section, we review the phantom dark energy model and derive the system of dynamical equations. In Sec. III, we find the cosmological fixed points and address the relation of the RG cutoff scale to cosmology. Finally, we close with a few concluding remarks in Sec. IV.

\section{Renormalization group}
RG is a powerful method to obtain the quantum effective action $\Gamma_k$ which integrates out all fluctuation modes with momentum larger than a certain cutoff $k$ and takes them into account through a modified dynamics for the remaining fluctuations with momentum smaller than $k$. This is implemented by means of a cutoff $R_k$ which, for a given $k$, suppresses the contributions of the low momentum modes, while integrates out the high-momentum modes in the path integral. Then the function $\Gamma_k$ defines an effective field theory valid near the scale $k$ and describes all quantum effects originating from the high-momentum modes. In particular, $\Gamma_k$ interpolates between the bare action $S_{\rm bare}\simeq \lim_{k\rightarrow \infty} \Gamma_k$ and the standard effective action $\Gamma \simeq \lim_{k\rightarrow 0} \Gamma_k$. The dependence of $\Gamma_k$ with RG scale obeys an exact functional differential equation \cite{Wetterich}
\begin{eqnarray}
k\partial_k\Gamma_k=\frac{1}{2}{\rm Tr} \left(\Gamma^{(2)}_k+R_k \right)^{-1} k\partial_k R_k,
\end{eqnarray}
which relates the change in $\Gamma_k$ with one-loop type integral over the full field-dependent cutoff propagator. Here,
$\Gamma^{(2)}_k[\phi]\equiv \delta^2 \Gamma_k/(\delta\phi \delta\phi)$, denotes the second functional derivative of this effective action with respect to the fields, which is the full inverse propagator at the scale $k$. The trace Tr denotes an integration over all momenta and summation over all fields.

To discuss the case where gravity and/or the matter sector display a non-trivial RG fixed point, we recall the RG equation for Newton coupling whose canonical dimension is $[G]=2-d$ in $d$ dimensions and hence negative for $d>2$. It is commonly believed that a negative mass dimension for the relevant coupling is responsible for the perturbative non-renormalisability of the theory. It is usually to introduce the renormalised coupling as $G_k=Z^{-1}_GG$, and the dimensionless coupling as $g(k)=k^2G_k$; the momentum scale $k$ denotes the renormalisation scale. The graviton wave function renormalisation factor $Z_G(k)$ is normalised as $Z_G(k_0)=1$ at $k=k_0$ with $G(k_0)$ given by Newton constant $G$. The graviton anomalous dimension $\alpha$ related to $Z_G(k)$ is given by $\alpha\equiv -k d\ln Z_G/dk$. Then the Callan每Symanzik equation for $g(k)$ reads
\begin{eqnarray}\label{RG}
\frac{d g(k)}{d \ln k}=(d-2+\alpha)g(k),
\end{eqnarray}
In general, the graviton anomalous dimension $\alpha$ is a function of all coupling of the theory including matter fields. The RG equation (\ref{RG}) displays two qualitatively different types of fixed point. The non-interacting (gaussian) fixed point corresponds to $g_*=0$ which also entails $\alpha=0$. In its vicinity, quantum effects are small and the gravitational couplings take their classical values $G_k\approx G$. In turn, equation (\ref{RG}) can display an interacting fixed point $g_*\neq 0$ in $d>2$ if the anomalous dimension takes the value $\alpha=2-d$. Consequently, at an interacting fixed point where $g_*\neq 0$, the anomalous dimension implies scaling: $G_k=g_*/k^{d-2}$. This fixed point scaling leads to a characteristic weakening of gravity at shortest distance $G_k\ll g$. Similarly, the effective potential $V_k$ achieves an RG fixed point provided it scales with its canonical dimension $V_k\propto k^4$, leading to $\beta=4$ (the definition of $\beta$ see the next section).

\section{Renormalization group improved phantom cosmology}
To relate RG ideals to cosmology, here we follow the approach adopted in \cite{Linder} (other methods see Refs. \cite{Hindmarsh,Reuter2004}), in which the couplings keep non-dynamical in the action and require only the Bianchi identity to hold with respect to the total covariant derivatives, simultaneously accounting for the spacetime dependence and the flow of the couplings under the RG. We assume a spatially flat Friedmann每Robertson每Walker (FRW) metric governed by Einstein gravity with matter and a minimally coupled phantom scalar field. The action of the phantom field minimally coupled to gravity is described by
\begin{eqnarray}
S=\int {\rm d}^4x\sqrt{-g}\left[\frac{1}{2}(\nabla \phi)^2-V(\phi) \right],
\end{eqnarray}
and the energy density and pressure density of phantom are given by, respectively,
\begin{eqnarray}
&& \rho_{\phi}=-\dot{\phi}^2/2+V(\phi),\\
&& p_{\phi}=-\dot{\phi}^2/2-V(\phi).
\end{eqnarray}
Fluctuations in a phantom field have a negative energy, therefore it may be possible for the vacuum to decay into a collection of positive-energy and negative energy
particles which may cause the model to be unstable. For example, in \cite{Carroll2003}, it has been shown that these models might be phenomenologically viable if it has been thought of as effective field theories valid only up to a certain momentum cutoff. But when higher order effects are considered, the models are unstable. This is the pressing issue of the known instability for phantom models \cite{Carroll2003,Carroll2005,Silvestri2009}. If the phantom theory is fundamental, the authors of reference \cite{Carroll2003} have calculated the decay rate of phantom particles into gravitons, and have found this decay rate would be infinite and valid up to arbitrarily high momenta, which renders the theory useless as a dark energy candidate. If considering instead the phantom theory to be an effective theory valid below a scale, with Lagrangian including operators of all possible dimensions which are suppressed by suitable powers of the cutoff scale, they have found that such higher order operators, even though they may be of very high order, can lead to unacceptably short lifetimes for phantom particles unless the cutoff scale is less than 100 MeV; so, it is difficult to reconcile with current accelerator experiments \cite{Carroll2003}. Since observations cannot rule out the possibility $w<-1$ \cite{Melchiorri2003, Gaztanaga2009}, and since we understand so little of dark energy, we should keep an open mind about the discussions of phantom. Besides the stability, the phantom model is also confronted with the problem of big rip which will be investigated here: can the big rip in phantom cosmology be avoid by taking into account effects coming from quantum gravity?

Because there is no explicit dependence of the coupling (including the gravitational coupling and the phantom potential) on the metric, the standard Friedmann equations arising from variation of the action with respect to the metric will be preserved
\begin{eqnarray}
\label{Fried}
&& H^2=\frac{K^2}{3}\left(\sum_i\rho_i- \frac{\dot{\phi}^2}{2}+V(\phi)\right),\\
&& \dot{H}=-\frac{K^2}{2}\left(\sum_i \gamma_i\rho_i- \dot{\phi}^2\right),
\end{eqnarray}
where $K^2=8\pi G$, $H=\dot{a}/a$ the Hubble parameter, and $\gamma_i=1+w_i$ with $w_i$ the $i$th EoS of barotropic fluid.

When we apply RG to general gravity, possibly coupled to matter fields, the gravitational and matter couplings turn into scale-dependent running coupling $G\rightarrow G_k$ and $V(\phi) \rightarrow V_k(\phi)$. Taking into account a piece from the possible time variation of the RG scale $k$, Einstein equation gives
\begin{eqnarray}
\label{bia}
0=(G_kT_k^{\mu\nu})_{;\nu}=(G_kT_k^{\mu\nu})_{,\nu}+G_k\Gamma^{\mu}_{\alpha\nu}T_k^{\alpha\nu}+G_k\Gamma^{\nu}_{\alpha\nu}T_k^{\mu\alpha}.
\end{eqnarray}
For $\mu=0$, one obtain the continuity equation
\begin{eqnarray}
\label{cone}
\dot{\rho_{k}}=-3H\rho_{k}\left[1+\frac{p_{k}}{\rho_{k}}+\frac{1}{3}\frac{d\ln k}{dN}\frac{\partial \ln(G_k\rho_{k})}{\partial \ln k} \right],
\end{eqnarray}
where $N\equiv \ln a(t)$. It is assumed that the RG scale parameter is a function of cosmological time, $k=k(t)$ (in our conventions $k=k(N)$). Equations (\ref{cone}) imply that the energy density $\rho_{k}$ is $k$-dependent which can takes the role of dark matter, dark energy, etc. See, for example, for dark matter component, it reads
\begin{eqnarray}
\dot{\rho_{\rm m}}=-3H\rho_{\rm m}\left[1+\frac{1}{3}\frac{d\ln k}{dN}\frac{\partial \ln G_k}{\partial \ln k} \right],
\end{eqnarray}
meaning the EoS of dark matter is non-zero in general, but rather $\frac{\alpha}{3}\frac{d\ln k}{dN}$. This will affect structure formation, which is beyond the scope of this article. However, the dark matter is irrelevant asymptotically in dark energy dominated epoch. The above equation is also a choice made in \cite{Linder}. In fact, it is perfectly possible to maintain the conventional conservation law for matter, as in Ref \cite{Hindmarsh} Eq. (20) is modified in that case.

In order to transform the cosmological equations into an autonomous dynamical system, we introduce the auxiliary variables (see, for example, Refs. \cite{Hindmarsh,Copeland1998,Hindmarsh1,Yang,Capozziello}):
\begin{eqnarray}\label{def}
x=\frac{K\dot{\phi}}{\sqrt{6}H},~~~~y=\frac{K\sqrt{V_k(\phi)}}{\sqrt{3}H},~~~~z=\frac{V_\phi}{KV_k},
\end{eqnarray}
where $V_\phi=dV_k/d\phi$. Using these variables we can straightforwardly obtain the density parameter and the EoS of phantom, respectively, as
\begin{eqnarray}
&& \Omega_\phi=-x^2+y^2,\\
&& w_\phi=\frac{x^2+y^2}{x^2-y^2}.
\end{eqnarray}
The self-autonomous system in terms of the auxiliary variable $x$ and $y$ are given by (similar to Eqs. (19)-(22) in Ref. \cite{Hindmarsh}):
\begin{eqnarray}
\label{aut}
&& x' = -3x(x^2+1)+\frac{\sqrt{6}}{2}z y^2+\frac{3}{2}x\sum_i \gamma_i \Omega_i +\frac{1}{2}x\alpha \frac{d\ln k}{dN} ,\\
\label{aut1}
&& y' = -3y\left(x^2-\frac{\sqrt{6}}{6}z x- \frac{1}{2}\sum_i \gamma_i \Omega_i\right)+\frac{1}{2}y(\alpha+\beta) \frac{d\ln k}{dN},\\
\label{aut2}
&& z'=\sqrt{6}x(\eta-z^2)+z(\zeta-\frac{1}{2}\alpha-\beta)\frac{d\ln k}{dN},\\
\label{aut3}
&& \Omega'_i=-3\Omega_i(2x^2-\gamma_i+\sum_i \gamma_i \Omega_i)+\alpha \Omega_i\frac{d\ln k}{dN},
\end{eqnarray}
where the prime denotes a derivative with respect to the logarithm of the scale factor, $\beta\equiv \partial \ln V_k/\partial \ln k$,
$\zeta\equiv \partial \ln V_{\phi}/\partial \ln k$, and $\eta \equiv V_{\phi\phi}/(K^2V_k)$. We do not consider the case where $\alpha= 0$ or $\beta= 0$ holds all the time, which means that $G_k$ or $V_k$ are independent on the scale $k$; but as $\alpha$ and $\beta$ evolve with time, we can have $\alpha(t_c)=0$ or $\beta(t_c)=0$ at certain time $t_c$.

With definitions (\ref{def}), the Friedmann constraint (\ref{Fried}) can be expressed as
\begin{eqnarray}
\label{cond3}
-x^2+y^2+ \sum_i \Omega_i=1.
\end{eqnarray}
For phantom dark energy, the continuity equation (\ref{cone}) gives
\begin{eqnarray}
\label{phcon}
\dot{\rho_{\phi}}=3H\dot{\phi}^2-\left[\alpha \left(-\frac{1}{2}\dot{\phi}^2+V_k \right)+\beta V_k \right]H\frac{d\ln k}{dN}.
\end{eqnarray}
Since $\phi$ does not explicitly depend on $k$, the variation of the action with respect to $\phi$ gives the Klein每Gordon equation, which must be consistent with the continuity equation obtained above. While in Klein每Gordon equation only the first term in the right hand side of equation (\ref{phcon}) appears, so the terms containing $d\ln k/dN$ must vanish for consistency of the theory.
This condition leads to two possibilities:
\begin{eqnarray}\label{cond1}
\frac{d\ln k}{dN}=0,
\end{eqnarray}
or
\begin{eqnarray}\label{cond2}
\alpha \left(-\frac{1}{2}\dot{\phi}^2+V_k \right)+\beta V_k=0.
\end{eqnarray}
Eq. (\ref{cond1}) implies that the RG flow freezes, while Eq. (\ref{cond2}) restricts arbitrary behavior of the RG flow for Einstein每Hilbert action. Under the condition (\ref{cond2}), we obtain
\begin{eqnarray}
\label{cond4}
(\alpha+\beta)y^2=\alpha x^2,
\end{eqnarray}
or, if $\alpha\neq 0$, $\beta\neq 0$ and $\alpha+\beta\neq 0$, equivalently the following equations hold:
\begin{eqnarray}
\label{cond5}
&& y^2=\frac{\alpha}{\alpha+\beta}x^2,\\
&& \Omega_\phi=-\frac{\beta}{\alpha+\beta}x^2,\\
&& w_\phi=1+2\frac{\alpha}{\beta}.
\end{eqnarray}
Constrained from (\ref{cond3}) and $x^2> 0$ (or $y^2> 0$), we have $\frac{\alpha}{\beta}< -1$, if $G$ and $V$ dependent on the RG scale. If $x=0$ or $y=0$, however, $\frac{\alpha}{\beta}< -1$ does not hold.

When Eq. (\ref{cond2}) is applied, the gravitational field equations, Bianchi identity and field evolution equation give a consistent framework within which to treat RG flow and cosmological dynamics together.

\section{Cosmological fixed points}
In this section, we discuss phantom cosmological fixed point solutions for RG improved cosmologies subjected to the continuity equation of phantom field.
To achieve a cosmological fixed point solution in the presence of a phantom scalar field, the RG scale $\ln k$ may or may not be forced to be constant with scale factor, depending on the value of $x$, $y$, $z$ and $\Omega_i$ in equations (\ref{aut}), (\ref{aut1}), (\ref{aut2}) and (\ref{aut3}). Hence two qualitatively different types of cosmological fixed point solution become available: a fixed point where the RG scale $\ln k$ become independent, and a fixed point where the RG scale continues to scale with $N$. For the case $d\ln k/dN=0$, the cosmological fixed points are formally the same as those of the standard phantom cosmologies; in other words, the classical fixed points appear as a particular solution for this case where no RG parameter dependence enters from the outset. Here, however, quantum corrections are present and incorporated through RG modifications in the variables $x$, $y$ and $z$.

We are interested in the evolution of the universe in late time. So we ignore the contribution of radiation. In addition, we are particularly interested in the effect on the cosmological behaviors from RG flow. To illustrate this point, we consider the exponential potential $V=V_0\exp(\lambda K\phi)$ for simplicity with $\lambda$ a dimensionless non-zero constant, which have been considered in literatures before. In this case, we have $z=\lambda$, $\eta=\lambda^2$, $\beta=\frac{1}{2}\lambda\phi K\alpha$ and $\zeta=\frac{1}{2}\alpha+\beta$. The cosmological fixed points of the autonomous system are obtained by setting the left hand sides of the equations to zero, namely letting $x'=y'=z'=\Omega'_i=0$, and taking into account Eqs. (\ref{cond3}) and (\ref{cond4}). We exhibit cosmological fixed solutions in Tab. \ref{crit}.

Considering $x_{\rm c0, c1}=0$ for points $P_{\rm c0}$ and $P_{\rm c1}$, we have $y_{\rm c0, c1}=0$ and $\Omega_{\rm m}=1$, a universe dominated by dark matter. $z_{\rm c0, c1}=\lambda\neq0$ and $w_{\rm \phi}$ can not be determined, and if taking $\alpha=0$, we also can not determine $\frac{d\ln k}{dN}$; see in Tab. \ref{crit} ($P_{\rm c0}$). Of course, if take $\frac{d\ln k}{dN}=0$ firstly, $\alpha$ can't be determined, too, see in Tab. \ref{crit} ($P_{\rm c1}$). We note, however, that the RG flow can reach the asymptotically safe UV fixed point in the latter case ($\frac{d\ln k}{dN}=0$). Because at the fixed point of RG flow, we have $\alpha=-2$ and $\beta=4$, and find cosmological fixed point is $\Omega_{\rm m}=1$, $x=y=0$, and $z=\lambda\neq 0$, which is just a special case of $P_{\rm c1}$. It is a interesting result that the universe and the RG flow can simultaneously reach their fixed point, and the universe is completely dominated by dark matter, free from the big rip. Unfortunately, however, we can not determine the stability of this point, because the determinant of the matrix of the linearized perturbation is zero. On taking other potentials in which the fixed point can be $x_{\rm c}=0$, $y_{\rm c}=0$, $z_{\rm c}=0$ and $\Omega_{\rm m}=1$, with $\alpha=-2$ and $\beta=4$, we can argue that it is an attractor in the future for $0<\eta<\sqrt{6}/2$.

%%%%%%%%%%%%%%%%%%%%%%%%%%%%%%%%%%%
\begin{table*}
\begin{center}
\begin{tabular}{|c|c|c|c|c|c|c|c|c|}
  \hline
  Case & $\Omega_{\rm m}$ & $x^2$ & $y^2$ & $z$ & $w_{\phi}$ & $\frac{d\ln k}{dN}$ & Existence & Type \\\hline
  $P_0$ & $1$ & $0$ & $0$ & $\lambda$ & $w_{\phi}$ & $\frac{d\ln k}{dN}$ & $\alpha=0$ & DM dominated \\\hline
  $P_1$ & $1$ & $0$ & $0$ & $\lambda$ & $w_{\phi}$ & $0$ & $\alpha\neq 0$ & DM dominated \\\hline
  $P_{2,3}$ & $0$ & $\frac{1}{6}\lambda^2$ & $1+\frac{1}{6}\lambda^2$ & $\lambda$ & $-1-\frac{1}{3}\lambda^2$ & $\frac{d\ln k}{dN}$ & $\alpha=\beta=0$ & mixed \\\hline
  $P_{4,5}$ & $0<\Omega_{\rm m}<1$ & $\frac{1}{2}(1-\Omega_{\rm m})$ & $\frac{3}{2}(1-\Omega_{\rm m})$ & $z^2=\frac{3(1-2\Omega_{\rm m})^2}{1-\Omega_{\rm m}}$ & $-2$ & $\frac{d\ln k}{dN}$ & $\alpha=\beta=0$ & scaling \\\hline
  $P_{6,7}$ & $0$ & $-(1+\frac{\alpha}{\beta})$ & $-\frac{\alpha}{\beta}$ & $\lambda$ & $1+2\frac{\alpha}{\beta}$ & $-\frac{6}{\beta}-\frac{\sqrt{6}}{\alpha+\beta}\lambda x$ & $\alpha\neq0, \beta\neq0, \alpha\neq -\beta$ & mixed \\\hline
  \end{tabular}
\end{center}
\caption{\label{crit} Renormalisation group improved cosmological fixed points for Einstein gravity with a phantom field
 and dark matter, including the values of related cosmological parameters.}
\end{table*}

Taking $\Omega_{\rm m}=0$ and $\alpha= \beta= 0$ for points $P_{\rm c2}$ and $P_{\rm c3}$, we get $x^2_{\rm c2, c3}=\frac{1}{6}\lambda^2$ and $y^2_{\rm c2, c3}=1+\frac{1}{6}\lambda^2$. $z_{\rm c2, c3}=\lambda\neq 0$ and $\frac{d\ln k}{dN}$ can not be determined, see in Tab. \ref{crit} ($P_{\rm c2, c3}$). The universe is dominated by phantom dark energy with $w_{\rm \phi}=-1-\frac{1}{3}\lambda^2$, implying that the big rip can not be avoided.

If $\alpha=\beta= 0$, but $\Omega_{\rm m}\neq 0$ and $\Omega_{\rm m}\neq 1$ for points $P_{\rm c4}$ and $P_{\rm c5}$, it is easy to find that $x^2_{\rm c4, c5}=\frac{1}{2}(1-\Omega_{\rm m})$, $y^2_{\rm c4, c5}=\frac{3}{2}(1-\Omega_{\rm m})$ and $z^2_{\rm c4, c5}=\lambda^2=\frac{3(1-2\Omega_{\rm m})^2}{1-\Omega_{\rm m}}$. The evolution of RG flow , $\frac{d\ln k}{dN}$, can not be determined. A universe is partially occupied by phantom dark energy with $w_{\rm \phi}=-2$, see in Tab. \ref{crit} ($P_{\rm c4, c5}$), and its fate dependents on the values of $\Omega_{\rm m}$ which can not be determined by the cosmological evolution equations.

For points $P_{\rm c6}$ and $P_{\rm c7}$, taking $\Omega_{\rm m}=0$ and $\alpha\neq0$, $\beta\neq 0$ and $\alpha\neq\beta$, the solutions are $x^2_{\rm c6, c7}=-(1+\frac{\alpha}{\beta})$, $y_{\rm c6, c7}=\sqrt{-\frac{\alpha}{\beta}}$ and $z_{\rm c6, c7}=\lambda$, see in Tab. \ref{crit} ($P_{\rm c6, c7}$). The RG scale parameter evolves as $\frac{d\ln k}{dN}=-\frac{6}{\beta}-\frac{\sqrt{6}}{\alpha+\beta}\lambda x_{\rm c6, c7}$. The universe is completely dominated by phantom dark energy with $w_{\rm \phi}=1+2\frac{\alpha}{\beta}<-1$. In this case, although the effect of RG flow presents, the big rip still cannot be avoided.

\section{Conclusions and discussions}
In this paper we have examined the application of renormalisation group ideals to phantom cosmology at late time. Keeping the form of the classical equations of motion invariant, we relate RG scale evolution to cosmological evolution. We have written down the RG improved cosmological evolution equations with running gravitational and matter couplings for arbitrary potential, related the RG scale via the quantities $\alpha$, $\beta$ and $\zeta$. We have obtained RG improved cosmological fixed points and the corresponding cosmological parameters, such as $\Omega_{\rm m}$ and $w_\phi$, under the exponential potential. We have found that the universe is dominated by dark matter at cosmological fixed point $P_{\rm c0, c1}$, meaning the big rip can be avoided. While at fixed point $P_{\rm c2, c3}$ and $P_{\rm c6, c7}$, the universe is dominated by phantom dark energy and will end in a big rip. At fixed point $P_{\rm c4, c5}$, the fate of the is universe dependents on the values of $\Omega_{\rm m}$ which we can not determine via cosmological evolution equations. When $\alpha=-2$ and $\beta=4$, the universe and the RG flow can simultaneously reach their fixed point, and the universe is free from the big rip. This point can also be an attractor in the future if we take potentials in which the fixed point can be $x_{\rm c}=0$, $y_{\rm c}=0$, $z_{\rm c}=0$, and $\Omega_{\rm m}=1$, with $0<\eta<\sqrt{6}/2$. This is an interesting results.

It would be valuable to solve the RG improved cosmological evolution equations for various other specific potentials and see whether big rip can be avoided. Other questions, such as how the universe evolves towards or away from fixed points and how the structure formation changes due to effect of the flow of the gravitational coupling, are worth to be discussed in future publications.

\begin{acknowledgments}
This study is supported in part by National Natural Science Foundation of China under Grant No. 11147028 the Hebei Provincial Natural Science Foundation of China under Grant No. A2011201147, and Research Fund for Doctoral
Programs of Hebei University under Grant No. 2009-155.
\end{acknowledgments}

\bibliography{apssamp}

\end{document}